\documentclass[12pt]{article}

\usepackage{amsmath}

\begin{document}


\def\a{\alpha}
\def\b{\beta}
\def\c{\varepsilon}
\def\d{\delta}
\def\e{\epsilon}
\def\f{\phi}
\def\g{\gamma}
\def\h{\theta}
\def\k{\kappa}
\def\l{\lambda}
\def\m{\mu}
\def\n{\nu}
\def\p{\psi}
\def\q{\partial}
\def\r{\rho}
\def\s{\sigma}
\def\t{\tau}
\def\u{\upsilon}
\def\v{\varphi}
\def\w{\omega}
\def\x{\xi}
\def\y{\eta}
\def\z{\zeta}
\def\D{\Delta}
\def\G{\Gamma}
\def\H{\Theta}
\def\L{\Lambda}
\def\F{\Phi}
\def\P{\Psi}
\def\S{\Sigma}
\def\V{\varPsi}

\def\o{\over}
\newcommand{\sla}[1]{#1 \llap{\, /}}

\newcommand{\beq}{\begin{eqnarray}}
\newcommand{\eeq}{\end{eqnarray}}
\newcommand{\gsim}{ \mathop{}_{\textstyle \sim}^{\textstyle >} }
\newcommand{\lsim}{ \mathop{}_{\textstyle \sim}^{\textstyle <} }
\newcommand{\vev}[1]{ \left\langle {#1} \right\rangle }
\newcommand{\bra}[1]{ \langle {#1} | }
\newcommand{\ket}[1]{ | {#1} \rangle }

\newcommand{\EV}{ {\rm eV} }
\newcommand{\KEV}{ {\rm keV} }
\newcommand{\MEV}{ {\rm MeV} }
\newcommand{\GEV}{ {\rm GeV} }
\newcommand{\TEV}{ {\rm TeV} }


\baselineskip 0.7cm

\begin{titlepage}

\begin{flushright}
UT-03-05 \\
\end{flushright}

\vskip 1.35cm
\begin{center}
{\large \bf
Warped QCD without the Strong CP Problem
}
\vskip 1.2cm
Akito~Fukunaga and Izawa K.-I.
\vskip 0.4cm

{\it Department of Physics, University of Tokyo,\\
     Tokyo 113-0033, Japan}

\vskip 1.5cm

\abstract{
QCD in a five-dimensional sliced AdS bulk with chiral extra-quarks
on the boundaries is generically free from the strong CP problem.
Accidental axial symmetry is naturally present
except for suppressed breaking interactions,
which plays a role of the Peccei-Quinn symmetry
to make the strong CP phase sufficiently small.
Breaking suppression and enhancement due to AdS warping are considered
in addition to naive boundary separation effects.
}
\end{center}
\end{titlepage}

\setcounter{page}{2}

\section{Introduction}

The standard model of elementary particles
as an effective theory including gravity
has two apparent fine-tuning problems which
are hard to be undertaken directly by additional (gauge) symmetries:
the cosmological constant and the strong CP
\cite{KC}%
\footnote{For some recent attempts, see Ref.\cite{Che}.}
problems.
The presence of extra dimensions might serve as
an alternative to symmetry which naturally
affects such fine-tuned parameters.

In this paper, following a previous one
\cite{IWY},
we proceed to consider
QCD in a five-dimensional sliced AdS bulk
\cite{RS}%
\footnote{In contrast, the previous paper
\cite{IWY}
deals with the case of flat bulk.}
with chiral extra-quarks on the boundaries
and confirm that it is generically free from 
the strong CP problem.
We have adopted the AdS bulk as a natural curved spacetime
background without the restriction to bulk flatness.

For definiteness,
let us suppose that there is a pair of extra-quarks
in addition to the standard-model quarks:
a left-handed colored fermion $\psi_L$ and 
a right-handed one $\psi_R$.
We assume an extra-dimensional space
which separates them
from each other along the extra dimension.
If the distance between them is sufficiently large,
the theory possesses an axial U$(1)_A$ symmetry
\begin{equation}
 \psi_L \rightarrow e^{i \alpha} \psi_L, \quad
 \psi_R \rightarrow e^{-i \alpha} \psi_R 
 \label{eq:chiral-transf}
\end{equation}
approximately,
whose breaking is suppressed
at a fundamental scale
\cite{IWY}. 
This accidental global symmetry, which is actually broken by 
a QCD anomaly, naturally plays a role of 
the Peccei-Quinn symmetry
\cite{PQ},
making the effective strong CP phase 
to be sufficiently small. 

The point is that the presence of such an approximate symmetry
is not an artificial requirement,
but a natural result stemming from the higher-dimensional geometry,
which might well be stable even against
possible quantum gravitational corrections.

\section{Bulk color gauge theory}

Let us consider the four-dimensional Minkowski spacetime $M_4$ 
along with one-dimensional extra-space $S^1$,
whose coordinate $y$ extends from $-l$ to $l$ (that is, two points
at $y=l$ and $y=-l$ are identified).
The SU(3)$_C$ gauge field is assumed to propagate on the whole
spacetime $M_4 \times S^1$ equipped with the AdS-slice metric
\beq
 ds^2=e^{-2\s}\y_{\m\n}dx^\m dx^\n-dy^2; \quad \s=k|y|,
 \label{ds2y}
\eeq
where $\mu, \n =0,\cdots,3$ and $k$ denotes a positive or negative
constant which determines the AdS curvature.

The action of the five-dimensional gauge field is given by
\beq
 S_A=\int d^4x \int _{-l}^l dy \, e^{-4\s} \frac{M_*}{4g^2(y)}
 {\rm tr}(F_{MN}F^{MN})
 +\int h(y) \,{\rm tr}(AF^2-\frac{1}{2} A^3F+\frac{1}{10}A^5)
 \label{5sa}
\eeq
where $F_{MN} = \partial_M A_N - \partial_N A_M + [A_M , A_N ]$
$(M,N=0,\cdots,4; x^4 = y)$,
$A=A_Mdx^M$, and $F=dA+A^2 = (1/2)F_{MN}dx^M dx^N$.
Here, $M_*$ is supposed to be a cutoff scale
in the higher-dimensional theory
and $g(y)$ and $h(y)$ are
gauge and Chern-Simons coupling functions, respectively.

Kaluza-Klein reduction to the four-dimensional spacetime, however, yields 
a massless color-octet scalar which is undesirable
in the low-energy spectrum.
Hence we consider an $S^1/{\bf Z}_2$ orbifold instead of the $S^1$.
The five-dimensional gauge field $A_M(x,y)$ is now under a constraint
\begin{equation}
 A_\mu(x,y)=A_\mu(x,-y), \quad 
 A_4(x,y)=-A_4(x,-y),
\end{equation}
which eliminates the scalar zero mode and
only yields a vector field at low energies.

In order to define the theory on the orbifold consistently, 
the action Eq.(\ref{5sa}) on the $S^1$ should be 
invariant under the ${\bf Z}_2$ transformation.
This invariance is achieved as long as $g(y)=g(-y)$ and $h(y)=-h(-y)$.
Note that the background metric Eq.(\ref{ds2y}) itself
has been chosen to be invariant
under the ${\bf Z}_2$ transformation.
We take the $g(y)$ to be $y$-independent
and the $h(y)$ as
\begin{equation}
 h(y)=c {y \o |y|},
\end{equation}
where c is a constant to be determined in the next section.%
\footnote{From a five-dimensional perspective,
eigenstates corresponding to massive Kaluza-Klein modes
have energies larger than $l^{-1}$,
which are to be integrated out in the reduction process.
Note that their four-dimensional apparent masses depend on the
location of each mode in the extra dimension
and could appear as a modified CFT on a boundary
\cite{Cha}.
We assume that
the effects of the modified CFT to the four-dimensional QCD
be phenomenologically viable,
though they are not fully understood due to their nonperturbative
nature (see also the localization arguments cited in the Discussion).
}
The gauge symmetries are left unbroken in the bulk:
the infinitesimal SU(3)$_C$ gauge transformation parameter is
restricted to satisfy $\c(x,y)=\c(x,-y)$.

\section{Boundary extra-quarks}

There are two fixed points in the $S^1/{\bf Z}_2$ orbifold: 
$y=0$ and $y=l$.
Let us put chiral extra-quarks
on the fixed-point boundaries:%
\footnote{
The standard-model quarks (and leptons)
are assumed to be on the fixed-point boundary at $y=0$.
We also include QCD $\theta$ and (possibly dominant)
Yang-Mills terms implicitly on the boundary 3-brane.}
a left-handed extra-quark $\psi_L$ at $y=0$
and a right-handed one $\psi_R$ at $y=l$.
The action of the split extra-quarks contains%
\footnote{
In fact, the extra-quarks may be directly connected through
superheavy modes (besides gauge interactions)
with masses of order $m$ $(\sim M_*)$ in the higher dimensions.
This effect is exponentially suppressed when $ml$
is large enough, which we estimate in the Appendix.}
\beq
 S_\p&=&\int_{y=0}d^4x \, \bar\p_Li{D\llap{/\,}}\p_L+\int_{y=l}d^4x \,
 e^{-3\s}\bar\p_Ri{D\llap{/\,}}\p_R \nonumber\\
 &=&\int_{y=0}d^4x \, \bar\p_Li{D\llap{/\,}}\p_L+\int_{y=l}d^4x \,
 \overline{\tilde\p}_R i{D\llap{/\,}}\tilde\p_R,
 \label{5sp}
\eeq
where we have defined the canonically normalized field
\beq
 \tilde\p_R = e^{-\frac 32kl}\p_R.
\eeq

Under an infinitesimal gauge transformation, this fermionic sector
provides a gauge anomaly
\beq
 \d S_{eff} = {i \o 24\pi^2}\int_{y=0}
            {\rm tr}\!\left(\c d(AdA+{1 \o 2}A^3) \right)
            - {i \o 24\pi^2}\int_{y=l}
            {\rm tr}\!\left(\c d(AdA+{1 \o 2}A^3) \right)
\eeq
due to its chirality,
though the fermion content is vector-like
from a four-dimensional perspective.

On the other hand,
the bulk action yields
\begin{eqnarray}
 \d S_A &=& \int h(y) \,{\rm tr}\!\left( (d \c) d(AdA+{1 \o 2}A^3) \right)
 \nonumber
 \\   
        &=& -\int (d h(y)) \,{\rm tr}\!\left( \c d(AdA+{1 \o 2}A^3) \right)
\end{eqnarray}
under the gauge transformation.
Gauge anomaly cancellation with the fermionic sector implies
\begin{equation}
 c = {i \o 48\pi^2}.
\end{equation}

\section{Anomalous quasi-symmetry}

The gauge-invariant theory is given by the total action
$S = S_A + S_\p$.
Then, there is an approximate axial U$(1)_A$ symmetry given 
by Eq.(\ref{eq:chiral-transf}).

\subsection{Spontaneous breaking}

The extra-quarks should be decoupled from the low-energy spectrum to
escape from detection. 
Thus, we introduce hypercolor gauge interactions in the bulk
to confine the extra-quarks at high energies.
Such new gauge interactions would
simultaneously induce a chiral condensate 
$\vev{\psi_{L} \tilde\psi^{\dagger}_R}$
to break down the axial U$(1)_A$ symmetry
and provide a corresponding Nambu-Goldstone (NG) boson
called an axion
\cite{WW}.
Nonvanishing anomaly U$(1)_A$[SU(3)$_C$]$^2$ induces a potential
of the axion field.

We adopt SU(3)$_H$ as the hypercolor gauge group and assume that
the chiral fermions on each boundary transform as 
$\psi_{L}({\bf 3},{\bf 3}^*)$ and $\psi_{R}({\bf 3},{\bf 3}^*)$
under the SU(3)$_C \times $SU(3)$_H$ gauge group.%
\footnote{
Extensions to larger gauge groups
and fermion representations are straightforward,
which are touched upon at the end of this section.}
The SU(3)$_H$ interaction is supposed to be
confining at an intermediate scale $F_a$ $(< M_*)$ and develop a chiral
condensate $\vev{\psi_{L} \tilde\psi_{R}^{\dagger}} \simeq F^3_a$.
Note that the gauge anomalies due to the fermionic sector
can be canceled by bulk Chern-Simons terms
in a similar way as in the previous section.

SU(3)$_H$-charged particles are confined and
only massless NG bosons are left at low energies.
If one switches off the SU(3)$_C$ gauge interaction,
there is the U(3)$_L \times$U(3)$_R$ flavor symmetry that acts on $\psi_L$
and $\psi_{R}^\dagger$.
The flavor symmetry U(3)$_L \times$U(3)$_R$ is spontaneously broken down 
to a diagonal U(3) symmetry.
However, there is not the U(3)$_L \times$U(3)$_R$ symmetry actually,
since a diagonal SU(3) is gauged as the SU(3)$_C$ gauge group.
Thus, the NG bosons
due to such chiral symmetry breaking transform as ${\bf 3} \times
{\bf 3}^* = {\bf adj.}+{\bf 1}$ under the SU(3)$_C$.
Moreover, 
the adjoint-part of the NG bosons acquire masses due to 
the SU(3)$_C$ radiative corrections. 
What remains massless is only the color-singlet NG boson,
which corresponds to the axial U$(1)_A$ symmetry in
Eq.(\ref{eq:chiral-transf}). 

The axial symmetry discussed above, however, also has 
U$(1)_A$[SU(3)$_H$]$^2$ anomaly.
Therefore, the color-singlet NG boson
obtains a large mass and it cannot play a role
of the axion for the color SU(3)$_C$.
Hence we further introduce an additional pair of chiral fermions 
on each boundary:
$\chi_{L}({\bf 1},{\bf 3}^*)$ at $y=0$
and 
$\chi_{R}({\bf 1},{\bf 3}^*)$ at $y=l$.
The global symmetry is now U(4)$_{L} \times$ U(4)$_{R}$
if the SU(3)$_C$ gauge interaction is neglected.
The strong dynamics of the SU(3)$_H$ gauge group
lead to chiral symmetry breakdown
$\vev{\psi_L \tilde\psi_{R}^{\dagger}} \simeq F_a^3$ and 
$\vev{\chi_L \tilde\chi_{R}^{\dagger}} \simeq F_a^3$.
Two color-singlets
would remain massless if it were not for anomalies.
In this case, one of them does play a role of the axion
that makes the effective strong CP phase
to be sufficiently small.

\subsection{Explicit breaking}

The accidental chiral symmetry discussed above is broken 
by effective operators of the chiral fermions.
When the condensation
$\vev{\psi_L \psi_{R}^{\dagger}} \simeq e^{{3 \over 2}kl}F_a^3$
is less than $M_*^3$,
dominant breaking is expected to come from
the lowest dimension operators, which we concentrate on
in this subsection.

The operators involving both `$\psi_L$ or $\chi_L$' and
`$\psi_R$ or $\chi_R$' may be highly suppressed.
In view of an example
of the mediator interactions investigated in the Appendix,
the breaking term is estimated as
\beq
 \frac{e^{-2kl}}{e^{ml}-e^{-ml}} M_* (\p_L \p_R^\dagger + \p_R \p_L^\dagger),
\eeq
where $m$ denotes the mediator mass.
This results in an axion potential term
\beq
 V_{\rm bulk}(a)
 \simeq \frac{e^{-\frac{1}{2}kl} M_* F_a^3}{e^{ml}-e^{-ml}}
 f_{\rm bulk}({a}/{F_a}),
\eeq
where $f_{\rm bulk}(a/F_a)$ is a function of order unity,
whose minimum is generically 
different from that of the potential induced exclusively by the QCD effects.
This finally yields an effective QCD $\theta$ parameter of order
\beq
 \h_{\rm bulk} \simeq \frac{e^{-\frac{1}{2}kl}M_* F_a^3}{(e^{ml}
 -e^{-ml})\L_{QCD}^4}
 \label{thetabulk}
\eeq
in the case with sufficient suppression.

On the other hand,
the operators involving either `$\psi_L$ and $\chi_L$' or
`$\psi_R$ and $\chi_R$' are expected to be suppressed
only by powers of $1/M_*$.
Such operators also induce an additional potential of the axion, 
though this correction does not necessarily spoil
the Peccei-Quinn mechanism:
Axial symmetry breaking operators on each boundary may have coupling
coefficients of order one.
Such operators with the lowest mass dimension are given by
\begin{equation}
 \int_{y=0} d^4x \, \frac{1}{M_*^5} (\psi_L)^3 (\psi_L)^3 + 
 \int_{y=l} d^4x \, e^{-4\s}\frac{1}{M_*^5}
 (\psi_R^\dagger)^3 (\psi_R^\dagger)^3 + {\rm h.c.}
\end{equation}
Integration of heavy particles with masses of order 
$F_a$ due to the SU$(3)_H$ interaction
induces an additional potential of the axion field $a$ as 
\begin{equation}
 V_{\q}(a) \simeq \frac{F_a^{14}}{(e^{-{1 \over 2}kl}M_*)^{10}}
 f_{\q}({a}/{F_a}).
\end{equation}
The resulting shift in the QCD $\theta$ parameter
is expected to be of order
\begin{equation}
 {\theta_{\q}} \simeq {{F_a}^{14} \over
 {(e^{-{1 \over 2}kl}M_*)^{10}}\Lambda_{\rm QCD}^4}
 \label{thetaq}
\end{equation}
again in the case with sufficient suppression.
We note that the axial symmetry breaking operators
on each boundary can be made to have higher mass dimensions
if we adopt a larger hypercolor gauge group
instead of the SU$(3)_H$.

Combining with the expression for the gravitational
scale $M_G \simeq 10^{18}\GEV$ in four dimensions
\beq
 M_G^2 = \frac{M^3}{k} (1-e^{-2kl})
 \label{5relation}
\eeq
given by the one $M$ $(\sim M_*)$ in five dimensions
\cite{RS},
the above results restrict possible values of the parameters
to circumvent the strong CP problem.
For example, $\h_{\q} < 10^{-9}$
is realized for $|kl| \lsim 15$
when $F_a \simeq 10^{10}\GEV$ and $M_*^2 \simeq 2lM^3$.

\section{Discussion}

When $F_a$ is larger than $e^{-{1 \over 2}kl}M_*$,
the analysis based on an operator power expansion
(as in the previous section) does not seem reliable.
However, even in such a case, the framework of effective theory
implies that the potential energy coupled to the original 
five-dimensional metric (based on the proper time)
be less than the cutoff scale.
Hence, instead of Eq.(\ref{thetabulk}), we obtain
\beq
 \h_{\rm bulk} \simeq \frac{(e^{-\frac12kl}M_*)^4}{\L_{QCD}^4}
\eeq
as a conservative estimate. We also obtain
\beq
 {\theta_{\q}}
 \simeq {{(e^{-{1 \over 2}kl}M_*)}^{8} \over F_a^4 \Lambda_{QCD}^4}
 \simeq {{(e^{-{1 \over 2}kl}M_*)}^{4} \over F_a^4} \h_{\rm bulk}
 < \h_{\rm bulk}
\eeq
instead of Eq.(\ref{thetaq}).

For example,  $\h_{\rm bulk} < 10^{-9}$ with $M_* \simeq 10^{18}\GEV$
is realized for $kl \gsim 100$.
This result might suggest a possible dual role played by a common bulk
with spacetime inflationary background
which simultaneously achieves a tiny cosmological constant
in four dimensions for $kl \gsim 140$
\cite{Iza}.
Then, the quantum dynamics of gluons
should be localized (due to the Yang-Mills term
\cite{Dva}%
\footnote{Alternatively, yet higher-dimensional
\cite{Oda}
analogues
or an $y$-dependent
\cite{Keh}
gauge coupling $g(y)$ may be considered.})
at the $y=0$ boundary,
so that they would be only partly affected
by the background curvature.

\section*{Acknowledgments}

We would like to thank K.~Fujikawa, J.~Hisano, T.~Watari,
and T.~Yanagida for valuable comments and discussions.

\appendix

\section{Appendix}

This appendix deals with five-dimensional bulk mediator fermions
and their solvable mixing with boundary fermions.

The kinetic term for a bulk fermion $\P$ on the AdS-slice
orbifold is given by
\cite{Gro}
\beq
 S_\P=\int d^4x \int_{-l}^l dy \, e^{-4\s}
 \bar\P D \P; \quad
 D = ie^\s\sla\q-2\g_5\s^\prime+\g_5\q_y
 \label{fermiz2}
\eeq
with the restriction either $\P(x,-y) = + \g_5 \P(x,y)$
or $\P(x,-y) = - \g_5 \P(x,y)$.
Here, the spin connection including $\s'$ has been taken into account
and the prime denotes differentiation with respect to $y$.

Let us introduce two fermions with opposite chiralities
\beq
 & & \P_1(x,-y)=\g_5\P_1(x,y), \nonumber\\
 & & \P_2(x,-y)=-\g_5\P_2(x,y)
 \label{pop12}
\eeq
along with a mass term:
\beq
 S_\V=\int d^4x \int_{-l}^l dy \, e^{-4\s}\bar\V\left(
 \begin{array}{cc} D & m \\ m & D \end{array}\right)\V; \quad
 \V = \left(\begin{array}{c} \P_1\\
 \P_2\end{array}\right).
 \label{defD}
\eeq

Natural mixing with the boundary fermions $\p_L$ and $\p_R$
is expected through such terms of order one coefficients as
\beq
 \int_{y=0}d^4x \, M_*^{\frac12}(\bar\P_2 \p_L+\bar\p_L \P_2)
 +\int_{y=l}d^4x \, e^{-4\s} M_*^{\frac12}(\bar\P_1 \p_R+\bar\p_R \P_1),
\eeq
when $\V$ is properly charged.
Integrating out the bulk fermions $\V$, we obtain, among others,
U$(1)_A$-breaking nonderivative terms of the form
\beq
 & &\int d^4x \, {1 \over 2}
 \left(\frac{e^{(2k+m)|y|}}{e^{2ml}-1}
 -\frac{e^{(2k-m)|y|}}{e^{-2ml}-1}\right)_{y=l}
 e^{-4kl} M_* (\bar\p_R \p_L+\bar\p_L \p_R)
 \nonumber\\
 &=&\int d^4x \, \frac{e^{-2kl}}{e^{ml}-e^{-ml}}
 M_* (\bar\p_R \p_L+\bar\p_L \p_R)
 = \int d^4x \, \frac{e^{-\frac12kl}M_*}{e^{ml}-e^{-ml}}
 (\overline{\tilde \p}_R \p_L+\bar\p_L {\tilde \p}_R),
\eeq
which are exponentially suppressed for large values of $ml$.

\end{document}